\documentclass[twocolumn,english,aps,prb,twocolum,a4paper,superscriptaddress,natbib,bibnotes,amsmath,amssymb,floatfix,groupedaddress,footinbib]{revtex4-2}
\usepackage[colorlinks=true,citecolor=blue,linkcolor=blue]{hyperref}
\pdfoutput=1
\usepackage[addedmarkup=colored, authormarkupposition=left]{changes} 
\usepackage{soul}

\definechangesauthor[name={TJK}, color=red]{TJK}
\usepackage[utf8]{inputenc}
\usepackage[english]{babel}
\usepackage{amsmath,amsfonts,amssymb}
\usepackage[T1]{fontenc}
\usepackage{xurl}

\usepackage{amsmath}
\usepackage{siunitx}
\usepackage[version=4]{mhchem}
\usepackage{amsfonts}
\usepackage{amssymb}

\usepackage{epstopdf}
\usepackage{graphicx}
\graphicspath{{./Figures/}}

\begin{document}

\title{Foundry compatible, efficient wafer-scale manufacturing of ultra-low loss, \\high-density Si$_3$N$_4$ photonic integrated circuits}

\author{Xinru Ji$^{1,2}$, Rui Ning Wang$^{1,2}$, Yang Liu$^{1,2}$, Johann Riemensberger$^{1,2}$, Zheru Qiu$^{1,2}$, and Tobias J. Kippenberg$^{1,2,\dag}$}

\affiliation{
$^1$Institute of Physics, Swiss Federal Institute of Technology Lausanne (EPFL), CH-1015 Lausanne, Switzerland\\
$^2$Center for Quantum Science and Engineering,Swiss Federal Institute of Technology Lausanne (EPFL), CH-1015 Lausanne, Switzerland
}

\maketitle

\noindent\textbf{
Silicon nitride (Si$_3$N$_4$) photonic integrated circuits (PICs) have shown low linear loss, negligible nonlinear loss, and high power handling over traditional silicon photonics~\cite{blumenthal2018silicon}.
To achieve high-density photonic integration and high effective nonlinearity through tight optical confinement, thick (e.g. $> 600$~nm for 1550 nm operation) stoichiometric Si$_3$N$_4$ films are indispensable~\cite{levy2010cmos,brasch2016photonic,ye2021overcoming,riemensberger2022photonic}.
However, when using low-pressure chemical vapor deposition (LPCVD) to achieve high optical material transparency, Si$_3$N$_4$ films exhibit large tensile stress on the order of GPa~\cite{yang2018characteristic}.
Methods for crack prevention are therefore essential.
The photonic Damascene process has addressed this issue, attaining record low loss Si$_3$N$_4$ PICs~\cite{pfeiffer2016photonic,pfeiffer2018photonic,liu2021high}, but it lacks control of the waveguide height, which is crucial for creating wide structures such as arrayed waveguide gratings.
Conversely, precise waveguide dimension and ultra-low loss have been achieved with subtractive processing~\cite{ji2017ultra,ji2021methods,ye2019high}, but this method is not compatible with mass production due to the use of electron beam lithography. 
To date, an outstanding challenge is to attain both lithographic precision and ultra-low loss in high confinement Si$_3$N$_4$ PICs that are compatible with large-scale foundry manufacturing.
Here, we present a single-step deposited, DUV-based subtractive method for producing wafer-scale ultra-low loss Si$_3$N$_4$ PICs that harmonizes these necessities.
By employing deep etching of densely distributed, interconnected trenches into the substrate, we effectively mitigate the tensile stress in the Si$_3$N$_4$ layer, enabling direct deposition of thick films without cracking and substantially prolonged storage duration.  
%A secondary ion mass spectrometry (SIMS) analysis reveals that these deep trenches simultaneously serve as gettering centers for metal impurities, in particular copper, thereby reducing the absorption loss in Si$_3$N$_4$ waveguides.
Lastly, we identify ultraviolet (UV) radiation-induced damage that can be remedied through a rapid thermal annealing~\cite{neutens2018mitigation}.
Collectively, we develop ultra-low loss Si$_3$N$_4$ microresonators and 0.5~m-long spiral waveguides with losses down to 1.4~dB/m at 1550~nm with high production yield.
This work addresses the long-standing challenges toward scalable and cost-effective production of tightly confined, low-loss Si$_3$N$_4$ PICs as used for quantum photonics~\cite{elshaari2020hybrid,madsen2022quantum}, linear and nonlinear photonics~\cite{liu2020integrated,moss2013new}, photonic computing~\cite{feldmann2021parallel}, Erbium based on-chip amplifiers~\cite{liu2022photonic}, and narrow-linewidth lasers~\cite{xiang2021laser,liu2023fully}.} 

\pretolerance=5000
%%%%%%%%%%%%%%%%%%%%%%%%%%%%%%%%%%%%%
\section*{Introduction}
The advent of optical fibers has brought about a transformative impact on our society, offering a remarkably low propagation loss of 0.14~dB/km~\cite{tamura2017lowest} and an aggregated data rate up to 319~Tb/s at 3001~km~\cite{puttnam2021319}.
However, the growing bandwidth demands of data centers and disaggregated computing, along with the increasing complexity and scalability of neuromorphic photonics~\cite{shastri2021photonics,qian2020performing} and optical clocks~\cite{kitching2018chip}, underscore the need for compact, efficient, and cost-effective photonic systems.
Integrated photonics has emerged as a promising solution to these challenges, providing compact, high-performance devices that operate at lower power levels than traditional fiber-based optics or bulky free-space components~\cite{thomson2016roadmap}.
Numerous material platforms have been developed for integrated photonics such as silica~\cite{lee2012chemically}, silicon~\cite{jalali2006silicon,foster2006broad}, aluminum nitride (AlN)~\cite{xiong2012aluminum}, lithium niobate ($\mathrm{LiNbO_3}$)~\cite{zhu2021integrated}, lithium tantalate ($\mathrm{LiTaO_3}$)~\cite{wang2023lithium}, AlGaAs~\cite{chang2019low}, gallium phosphide (GaP)~\cite{wilson2020integrated}, and chalcogenide glass~\cite{zakery2003optical,li2014integrated}. 
Among these, silicon nitride (Si$_3$N$_4$) has garnered significant attention due to its unique combination of high power handling, space compatibility, high effective Kerr nonlinearity, and large transparency window covering the optical C-band.
Such low-loss Si$_3$N$_4$ photonic integrated circuits (PICs) have driven advancements in several areas, including large-scale optical networks for quantum photonics~\cite{elshaari2020hybrid,madsen2022quantum} and optical computing~\cite{feldmann2021parallel}, sensing~\cite{zhang2023photonic}, integrated lasers with fiber laser-level coherence and ultra-fast frequency tuning~\cite{jin2021hertz,xiang2021laser,lihachev2022low,siddharth2023hertz}, and, more recently, on-chip erbium-doped amplifiers~\cite{liu2022photonic} and lasers~\cite{liu2023fully}.
Equally important, the low-loss Si$_3$N$_4$ PICs with tight optical confinement have become the workhorse for integrated nonlinear optics, from which dissipative Kerr solitons (DKS) can be generated~\cite{herr2014temporal,brasch2016photonic}.
This forms the basis for emerging applications such as microcomb-based coherent optical communications~\cite{marin2017microresonator}, massive parallel LiDAR~\cite{riemensberger2020massively} and frequency metrology~\cite{udem2002optical}.
\begin{figure*}
     \centering
     \includegraphics[width=1\textwidth]{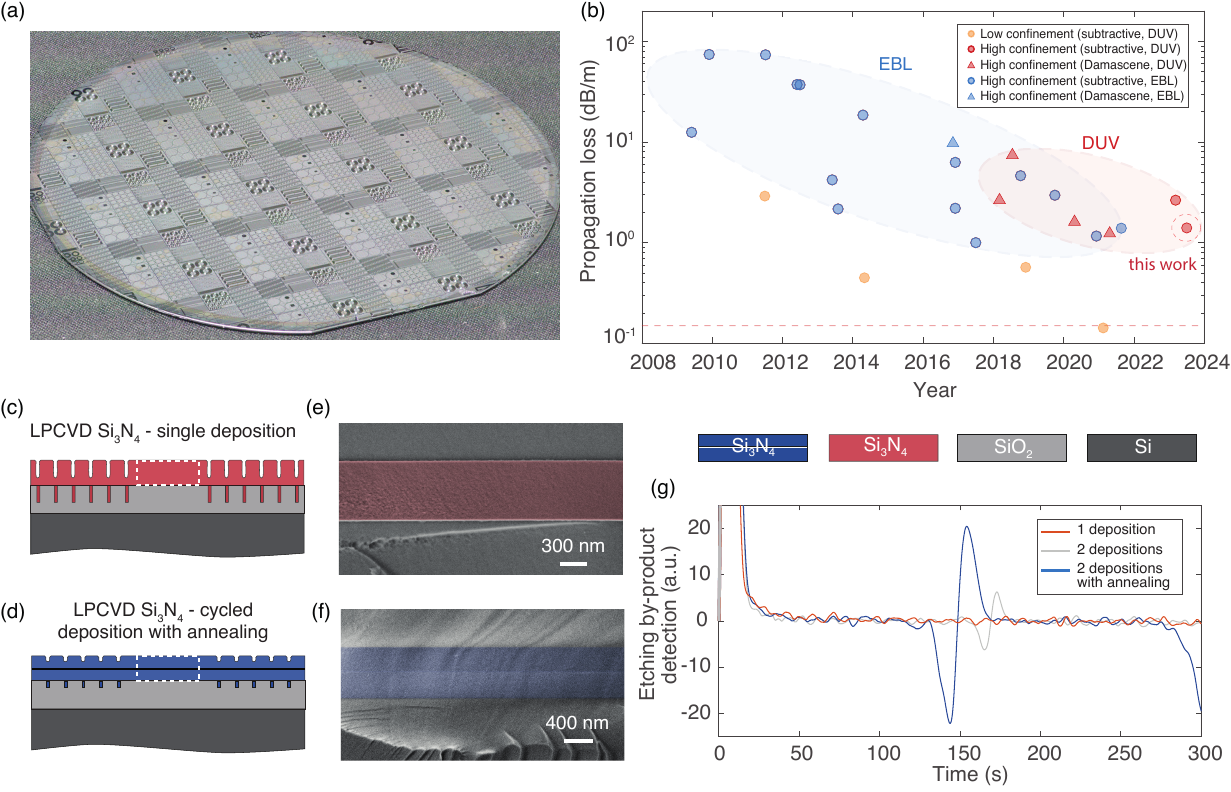}
    \caption{\textbf{Comparison of low-loss linear and nonlinear Si$_3$N$_4$ waveguide platforms via advanced fabrication methods.} 
    \textbf{(a)} A picture of a 4-inch wafer filled with high-density Si$_3$N$_4$ PICs, fabricated using DUV lithography and single-step LPCVD for subtractive processing.
    \textbf{(b)} The survey of Si$_3$N$_4$ waveguide loss evolution over the years. 
    The blue shaded elliptical area indicates prior work using EBL~\cite{gondarenko2009high,okawachi2011octave,levy2012high,riemensberger2012dispersion,luke2013overcoming,li2013vertical,pfeifle2014coherent,xuan2016high,ji2017ultra,stern2018battery,ye2019high,ji2021exploiting,ye2021overcoming,pfeiffer2016photonic}, while the red shaded elliptical area highlights those using DUV lithography~\cite{liu2018ultralow,pfeiffer2018ultra,liu2020photonic,liu2021high,ye2023foundry}.  
    The dashed red circle indicates the propagation loss of our device, the lowest recorded for the nonlinear Si$_3$N$_4$ waveguide, achieved with DUV stepper lithography and subtractive processing.
    The area under the red-dashed line indicates propagation loss under 0.15~dB/m, marking the absorption limit in tightly confined Si$_3$N$_4$ platforms~\cite{liu2021high}.
    \textbf{(c)(d)} Diagram illustrating the formation of thick Si$_3$N$_4$ films via single-step low-pressure chemical vapor deposition (LPCVD) and thermally cycled LPCVD techniques.
    \textbf{(e)} Scanning electron microscope (SEM) images of single-step LPCVD Si$_3$N$_4$ film and \textbf{(f)} cycled LPCVD Si$_3$N$_4$ film with intermediate annealing.
    \textbf{(g)} A comparative analysis of etching end-point graphs for Si$_3$N$_4$ films: single-step (red), thermal cycles (grey), and thermal cycles with intermediate annealing (blue). 
    Midpoints vary between the cycled films due to differences in layer thickness.}
     \label{fig:0}
\end{figure*}

For on-chip nonlinear applications, anomalous group velocity dispersion and tight optical confinement require waveguide thicknesses over 600~nm~\cite{levy2010cmos,brasch2016photonic}, which however leads to film cracking due to high tensile stress~\cite{yang2018characteristic}.
The photonic Damascene waveguide manufacturing process has emerged as a reliable high-yield method to overcome this issue, and can produce ultra-high quality factor ($Q>30$ million) and high confinement Si$_3$N$_4$ waveguides with thickness exceeding 900~nm~\cite{pfeiffer2016photonic,liu2021high}.  
In this process, waveguides are created using cycled depositions of LPCVD Si$_3$N$_4$ films (with a total thickness up to 1.1~$\mu$m) to reflown waveguide preforms with reduced sidewall roughness, followed by chemical-mechanical polishing to achieve sub-nm top surface roughness and reach the desired thickness.
Despite the increased complexity, the photonic Damascene process has enabled numerous breakthrough demonstrations, including system-level microcomb applications~\cite{riemensberger2020massively}, turnkey microcombs~\cite{shen2020integrated}, erbium-doped on-chip amplifiers~\cite{liu2022photonic} and net gain parametric amplifiers~\cite{riemensberger2022photonic}. 
However, this approach lacks precise control over waveguide dimensions due to height variations from etching and surface polishing processes, as well as unavoidable width changes caused by thermal preform reflow.
These dimensional changes lead to unpredictable variations in waveguide dispersion and the mode interaction between evanescently coupled waveguides, which consequently hinder the practical yield of usable devices such as microring resonators for DKS generation, power splitters, or wavelength-division multiplexed couplers.

The subtractive process that directly patterns and etches waveguides after Si$_3$N$_4$ film deposition, routinely offers precise control of waveguide dimensions and maintains high thickness uniformity across the wafer.
However, fabricating tightly confined optical waveguides using high-stress, thick Si$_3$N$_4$ films remains challenging due to the increased risk of wafer cracks and reduced production yield.
Alternative techniques, such as two-time deposition with cycled annealings for thick Si$_3$N$_4$ film formation, have been employed to alleviate large film stress~\cite{ji2021methods,ye2019high,el2019ultralow,ye2023foundry}, but increase fabrication costs, and add uncertainty to film uniformity. 
To date, the lowest loss tightly confined Si$_3$N$_4$ PICs fabricated with the subtractive method rely on electron beam lithography (EBL)~\cite{ji2017ultra,ye2019high,ye2021overcoming} which hinders scalable processing and limits dimension expansion for large-scale photonic circuits.
\begin{figure*}[t]
    \centering\includegraphics[width=1\textwidth]{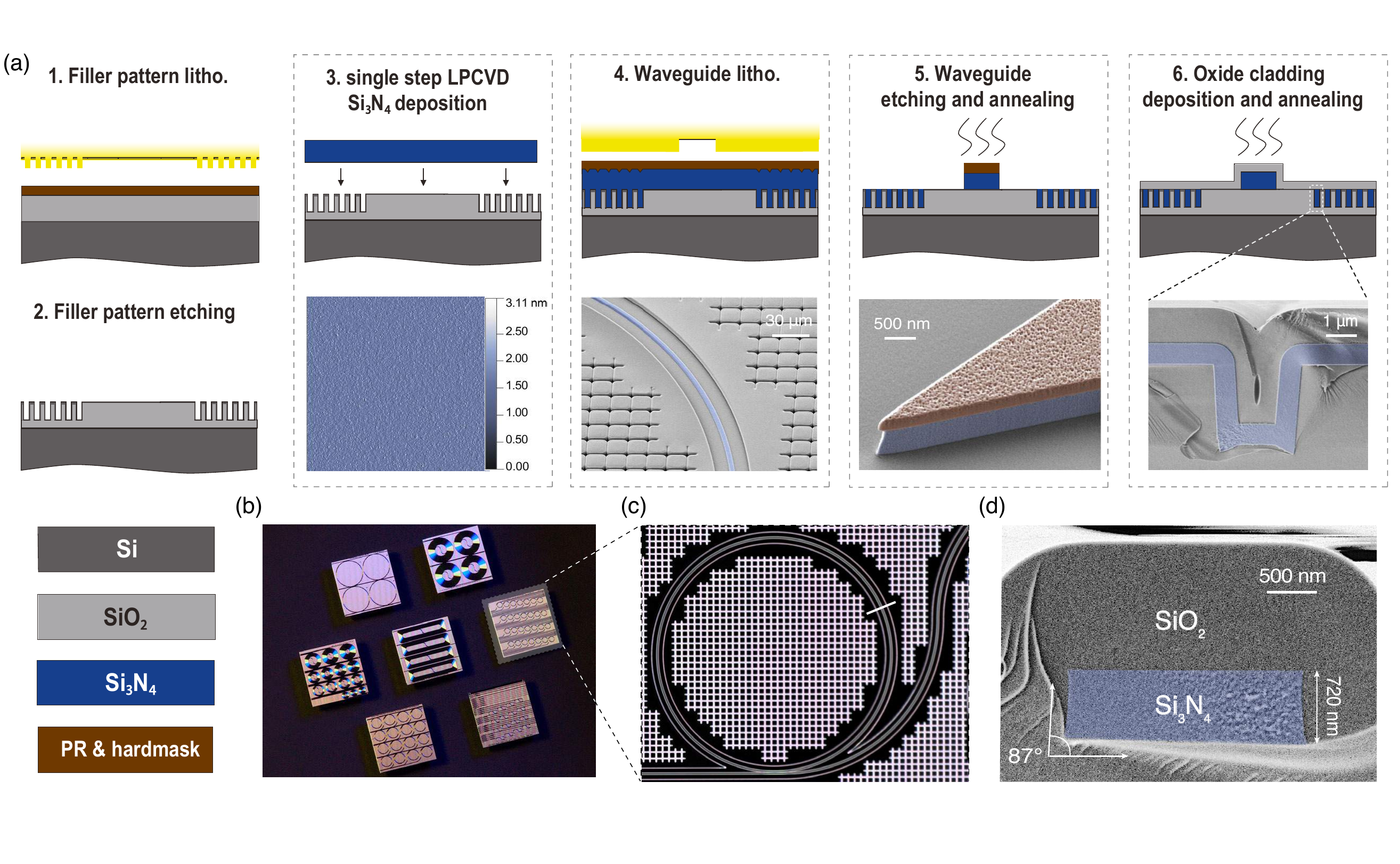}
    \caption{\textbf{Process flow for wafer-scale subtractive fabrication of ultra-low loss, tightly confining Si$_3$N$_4$ PICs with single-step deposition.}
     \textbf{(a)} The subtractive process used in this work includes deep ultraviolet (DUV) stepper lithography, preform dry etching, single-step LPCVD Si$_3$N$_4$ deposition, high-temperature annealing, and SiO$_2$ cladding deposition. 
%    The filler pattern trenches are etched to a depth of 3.5~$\mu$m, and a thick Si$_3$N$_4$ film (>700~nm) is deposited in a single layer subsequently. 
    A scanning electron microscope (SEM) image of the etched Si$_3$N$_4$ waveguide illustrates the smooth sidewall after step 5. The porous top material represents the SiO$_2$ etching hard mask, which is identical to the top cladding material. 
    Panel 5 displays the cross-section of filler patterns after Si$_3$N$_4$ and SiO$_2$ cladding deposition. 
    \textbf{(b)} A photograph of fabricated Si$_3$N$_4$ photonic chips featuring ring resonators with different free spectral ranges (FSR) and long spiral waveguides. All these chips have a footprint measuring 5$\times$5 mm$^2$. \textbf{(c)} A zoomed-in optical microscopic image of the highlighted chip in (b), featuring a microring resonator with a radius of 230~$\mu$m, a height of 720~nm, and a width of 2.6~$\mu$m, side-coupled with a bus waveguide. \textbf{(d)} An SEM image of the cross-section of the Si$_3$N$_4$ waveguide, showing a measured sidewall angle of 87$^\circ$.}
    \label{fig:1}
\end{figure*}
In this work, we achieve the combination of previously incompatible properties of  high-density,  ultra-low loss (down to 1.4~dB/m at 1550~nm) Si$_3$N$_4$ photonic integrated circuits, lithographic precision, and scalable wafer manufacturing (Fig.~\ref{fig:0}(a),(b)).
We present a novel, efficient single-step deposition subtractive process using deep ultraviolet (DUV) stepper lithography.
Our approach stands out for its efficiency and the prevention of cycling-induced film irregularities for the wafer-scale manufacturing of reliable photonic devices.
Notably, it also extends the storage life of thick LPCVD Si$_3$N$_4$ films, essential for mass production of photonic integrated circuits increasingly demanded for emerging applications (Supplementary Information Note 1).

\section*{Fabrication method}
We started with a 4~$\mathrm{\mu m}$ SiO$_2$ film grown by wet oxidation on 525~$\mu$m thick single-crystalline silicon wafers.
To release the large tensile stress in the Si$_3$N$_4$ film and prevent crack formation, we introduced interconnected filler patterns on the SiO$_2$ bottom layer.
Deep trenches were created by dry etching the patterns to a depth of 3.5~$\mu$m, enclosing areas with controlled stress and disrupting force accumulation. 
A scanning electron microscope (SEM) image shown in Figure~\ref{fig:1}(a) (Panel 4) highlights the structural features of the filler patterns.
Next, we employed a single-step low-pressure chemical vapor deposition (LPCVD) to precisely form the thick Si$_3$N$_4$ layer.
The as-deposited Si$_3$N$_4$ film exhibits a thickness variation of 1.2$\%$ (720~nm in total) and a smooth top surface characterized by an RMS roughness of 0.3~nm (Fig.~\ref{fig:1}(a)(Panel 3)). 
The differences between the LPCVD Si$_3$N$_4$ films grown with thermal cycles and a single-step deposition are presented in Fig.~\ref{fig:0}(c)-(g).
Notably, stacked films with intermediate annealing exhibit discernible boundaries (Fig.~\ref{fig:0}(f)). 
For Si$_3$N$_4$ films cycled without such annealing, no visible boundary is observed, but non-uniformities are revealed by an endpoint midway through the waveguide etching (Fig.~\ref{fig:0}(g), and Methods). 
For waveguide patterning, we used a KrF 248~nm deep-ultraviolet (DUV) stepper lithography with a resolution of 180~nm.
To date, the lowest reported losses in tightly confined Si$_3$N$_4$ PICs have involved EBL, which exhibits susceptibility to the environment and constrained throughput~\cite{ji2017ultra,ji2021exploiting,ye2021overcoming}. 
In contrast, DUV stepper lithography offers improved reliability and higher capacity~\cite{robinson2016materials}.
%%%
We then performed anisotropic dry etching based on a C$_x$F$_y$-chemistry to transfer the waveguide patterns onto the Si$_3$N$_4$ layer. 
To achieve smooth waveguide sidewalls, an LPCVD SiO$_2$ layer was deposited as an etching hard mask on top of the Si$_3$N$_4$ film. 
Oxygen was introduced during the waveguide etching process to remove CF polymer by-products of the etching reaction. 
The SEM image in Fig.~\ref{fig:1}(a)(Panel 5) shows a smooth waveguide sidewall, and the porous layer visible on top of the waveguide corresponds to the SiO$_2$ hard mask after etching. 
The waveguide cross-section (Fig.~\ref{fig:1}(d)) demonstrates a trapezoidal shape with a measured sidewall angle of 87$^{\circ}$.

The Si$_3$N$_4$ waveguides were subjected to a high-temperature annealing process at 1200~$^\circ \text{C}$ for 11~hours, which can remove excess H$_2$ and break the Si-H and N-H bonds within the Si$_3$N$_4$ waveguide core to reduce absorption loss (Supplementary Information Note 2 for detailed comparison of various annealing).
Subsequently, we employed LPCVD SiO$_2$, identical to the hard mask material, to clad the Si$_3$N$_4$ waveguides. 
This approach eliminates the need for additional steps to remove the hard mask, preserving the smoothness of the waveguide's top surface. 
The total thickness of the upper SiO$_2$ cladding is approximately 1.3~$\mu$m (Fig.~\ref{fig:1}(d)).
We subjected the SiO$_2$ cladding to the same high-temperature treatment (1200$^\circ \text{C}$, 11 hours) as the Si$_3$N$_4$ waveguides, to minimize absorption losses by disrupting O-H bonds and diffusing out H$_2$.
Additionally, this annealing process consolidates the SiO$_2$ film, leading to a reduction in surface RMS roughness (Supplementary Information Note 2).

\section*{Results}
\subsection*{Photonic integrated circuit characterization}
\begin{figure*}
    \centering
    \includegraphics[width=1\textwidth]{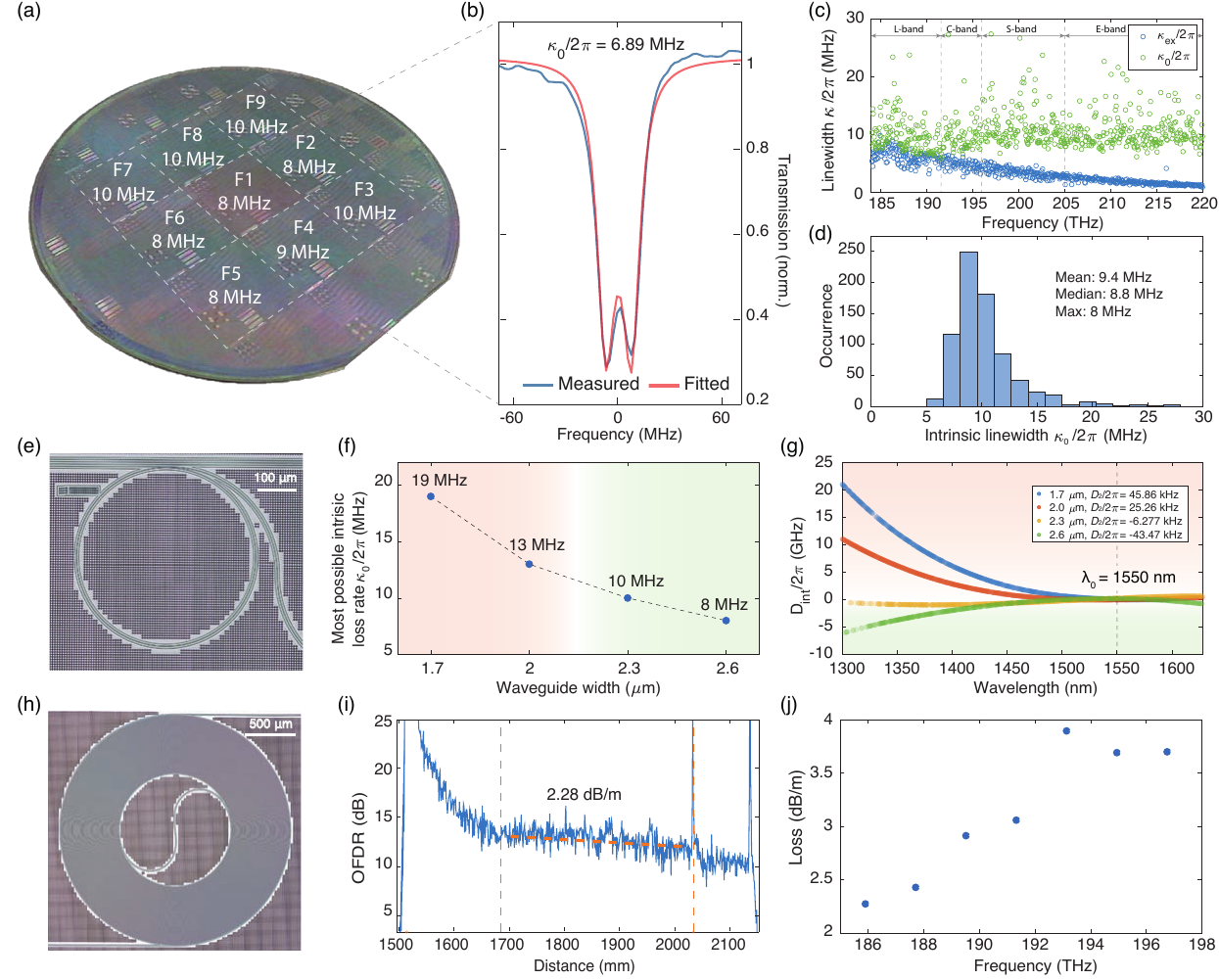}
     \caption{\textbf{Wafer-scale loss characterization of microring resonators and spiral waveguides.} \textbf{(a)} Wafer-scale measurement map of the most probable intrinsic loss $\kappa_0$/2$\pi$ for resonators with a 50~GHz free spectral range (FSR) and a waveguide width of 2.6~$\mu$m, fabricated with the subtractive process in Fig.~\ref{fig:1}. The measurement is performed using scanning laser spectroscopy. 
     \textbf{(b)} A zoomed-in view of a representative normalized resonance with an intrinsic linewidth of 6.89~MHz at 192.94~THz, corresponding to $Q\sim$~28 million. \textbf{(c)} Broadband measurement of the intrinsic linewidth $\kappa_0$/2$\pi$ and the external coupling rate $\kappa_{ex}$/2$\pi$ for a 50~GHz ring resonator with a 2.6~$\mu$m waveguide width in stepper field 1. \textbf{(d)} Histogram displaying the most probable value of $\kappa_0$/2$\pi$ obtained from the fitted data in \textbf{(c)}. 
     \textbf{(e)} A microscopic image of a 50~GHz ring resonator characterized in \textbf{(c)} and \textbf{(d)}.
     \textbf{(f)} Investigation of width-dependent intrinsic linewidth for 50~GHz resonators.
     \textbf{(g)} Analysis of integrated dispersion ($D_\text{int}/2\pi$) for resonators from \textbf{(f)}, highlighting regions of anomalous (red) and normal (green) dispersion.
     \textbf{(h)} A microscopic image of a 50-cm-long spiral waveguide. \textbf{(i)} Optical frequency domain reflectometry (OFDR) analysis of the spiral waveguide in \textbf{(h)}, using segmented Fourier transformation; waveguide dimensions are 2100~nm$\times$720~nm with a propagation loss of 2.28~dB/m.
     \textbf{(j)} Propagation loss extraction from OFDR traces at seven frequency points within the optical C-band.}
    \label{fig:2}
\end{figure*}

\begin{figure*}
     \centering
     \includegraphics[width=1\textwidth]{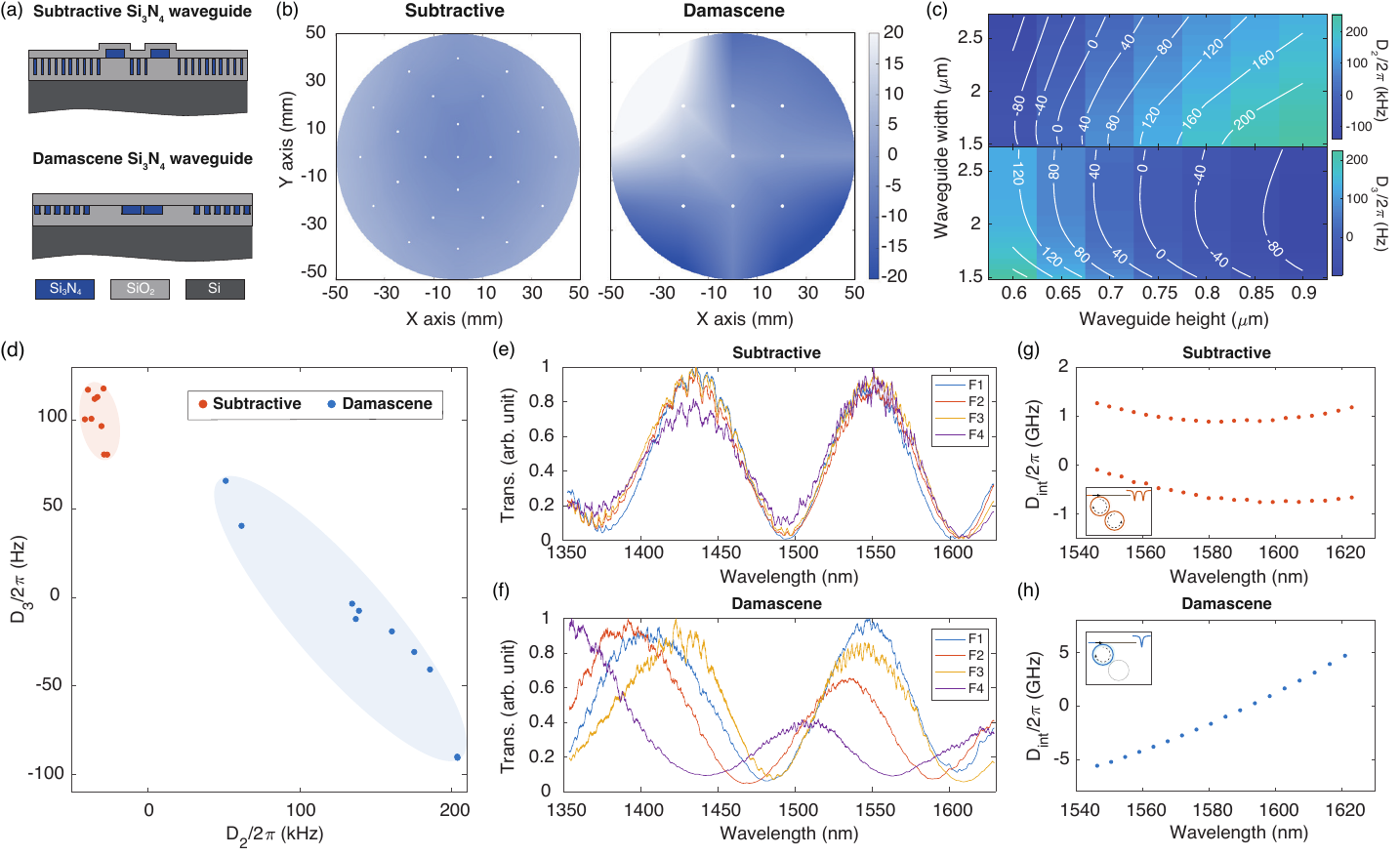}
    \caption{\textbf{Comparative analysis of Si$_3$N$_4$ waveguides via photonic Damascene and subtractive processes.}
\textbf{(a)} Schematic of Si$_3$N$_4$ waveguide structures using both fabrication techniques.
\textbf{(b)} Si$_3$N$_4$ waveguide height variation map for both processes, centered at average values of 815~nm for the subtractive wafer and 690~nm for the Damascene wafer.
\textbf{(c)} Simulation of group velocity dispersion ($\mathrm{D}_2/2\pi$) and higher-order dispersion ($\mathrm{D}3/2\pi$) for the TE${00}$ mode in a 100~GHz FSR Si$_3$N$_4$ microresonator, depicted as functions of waveguide dimensions.
\textbf{(d)} Experimental dispersion ($\mathrm{D}_2/2\pi$, $\mathrm{D}_3/2\pi$) in Si$_3$N$_4$ microresonators, contrasting both fabrication methods (subtractive in red,  Damascene in blue).
\textbf{(e-f)} Transmission characteristics of wavelength-division multiplexed couplers, comparing both methods.
\textbf{(g-h)} Integrated dispersion in coupled resonator systems via each process, with subtractive (red) and Damascene (blue) highlighted.}
     \label{fig:dispersion}
\end{figure*}

We investigated the waveguide loss across the central nine stepper fields by performing wafer-scale characterization of the resonance linewidths of the microring resonators (Fig.~\ref{fig:2}(a),(b)), using frequency-comb-assisted broadband (1260 to 1630~nm) laser spectroscopy~\cite{del2009frequency,liu2016frequency} at transverse electric (TE) polarization. 
By fitting each resonance~\cite{li2013unified}, we extracted the wavelength-dependent intrinsic loss rate $\kappa_0/2\pi$, and the bus-waveguide-to-resonator coupling rate $\kappa_\text{ex}/2\pi$ (Fig.~\ref{fig:2}(c)).
The loss histogram (Fig.~\ref{fig:2}(d)) shows the most probable value of $\kappa_0/2\pi=$ 8~MHz for devices in field 1, corresponding to a linear loss $\alpha_L$ of 1.4~dB/m and an intrinsic quality factor ($Q$) of $24 \times 10^6$ at 1550~nm. 
A minimum intrinsic linewidth of $\kappa_0/2\pi = 6.89$~MHz is observed at 1553.8~nm (Fig.~\ref{fig:2}(b)), corresponding to $\alpha_L = $1.32~dB/m and $Q = 28 \times 10^6$.
We next characterized the loss dependence on the waveguide width ranging from 1.7 to 2.6~$\mu$m of resonators with a free spectral range (FSR) of 50~GHz (Fig.~\ref{fig:2}(f)).
A clear inverse correlation between $\kappa_0/2\pi$ and the width arises since the interaction between optical modes and sidewalls diminishes as the waveguide cross-section expands, which aligns with the simulation based on the equivalent current model~\cite{payne1994theoretical} (Supplementary Information Note 4).
%%%%%%

To demonstrate the versatility of our Si$_3$N$_4$ platform for flexible dispersion engineering, we accessed both anomalous and normal dispersion in microresonators solely by varying the waveguide width.
Fig.~\ref{fig:2}(g) shows the measured integrated microresonator dispersion fitted by $\mathrm{D_\text{int}(\mu)=\omega_{\mu}-\omega_0-D_1\mu=D_2\mu^2/2+D_3\mu^3/6+D_4\mu^4/24}$ where $\omega_{\mu}/2\pi$ is the $\mu$-th resonance frequency relative to the reference resonance frequency $\omega_0/2\pi$.
$\mathrm{D}_1/2\pi$ corresponds to the microresonator FSR, $\mathrm{D}_2/2\pi$ is group velocity dispersion (GVD), and $\mathrm{D}_3$ and $\mathrm{D}_4$ are higher-order dispersion terms.
The region shaded in red represents the anomalous-GVD, which is accessed when the waveguide width is smaller than 2.3~$\mu$m at 720~nm height in these resonators with 50~GHz FSR.
This waveguide dimension is well suited for soliton microcomb generation and parametric amplification applications.

%%%% spiral OFDR loss %%%%
We verified that the low loss can be sustained in long spiral waveguides---an important building block for implementing optical delay lines, supercontinuum generation, and large scale photonic circuits --- which are extremely challenging for fabrication using electron beam lithography.
We evaluated the propagation loss of 0.5-meter-long spiral waveguides fabricated on the same wafer (Fig.~\ref{fig:2}(h)), using the optical frequency-domain reflectometry (OFDR) technique.  
The spiral waveguide has a footprint of 3.829~mm$^2$, a cross-section of 2100~nm$\times$720~nm, and a separation distance of 3~$\mu$m between adjacent waveguides.
We determined the propagation losses by linearly fitting the distance-dependent optical back reflectivity (represented by the dotted orange line in Fig.~\ref{fig:2}(i)), while disregarding a sporadic scattering feature. 
By conducting segmented Fourier transform on the fitted linear OFDR traces, we derived propagation losses ranging from 2.28 to 3.88~dB/m within the optical C-band (Fig.~\ref{fig:2}(j)).
These losses exhibit an upward trend with measurement frequency, indicative of a dominant contribution from Rayleigh scattering that quadratically increases as the wavelength decreases.

\subsection*{High waveguide thickness uniformity for dispersion and dimension control}
Our approach that combines single-step film deposition and subtractive process enables greatly improved waveguide thickness uniformity for consistent and precise dimension control of waveguide dispersion and optical evanescent coupling, which are critical to reliably implementing key building blocks such as dispersion-engineered microring resonators, directional couplers, and coupled resonators. 
In contrast, this was previously not achievable in low-loss Si$_3$N$_4$ photonic chips fabricated using the photonic Damascene process (Fig.~\ref{fig:dispersion}(a)). 
Our process demonstrates a minimal 9~nm variation (1.1$\%$) across a 4-inch wafer, one order of magnitude smaller than the 42~nm variation (6.1$\%$) observed with the photonic Damascene process (Fig.~\ref{fig:dispersion}(b)).
This outperforming advantage lies in the elimination of preform etching and chemical mechanical polishing in our approach, which are unavoidable in the photonic Damascene process.
Numerical simulations indicate that thickness variations of 40~nm can alter the group velocity dispersion ($\mathrm{D}_2/2\pi$) by up to 40~kHz (Fig.~\ref{fig:dispersion}(c)).
In experiment, measured dispersion variation exceeds 150~kHz in $\mathrm{D}_2/2\pi$ and 160~Hz in $\mathrm{D}_3/2\pi$, which implies the additional uncertainty of the waveguide width in the photonic Damascene process (Fig.\ref{fig:dispersion}(d)).
Conversely, our approach results in consistent and tight dispersion control, with a variation of 15~kHz in $\mathrm{D}_2/2\pi$ and 30~Hz in $\mathrm{D}_3/2\pi$.

To verify our approach's reliable and consistent fabrication of dimension-sensitive photonic components, we fabricated two high-density Si$_3$N$_4$ wafers with functional devices including wavelength-division multiplexed (WDM) couplers and coupled-resonator systems, using our process and the photonic Damascene process, respectively.
The WDM coupler based on optical evanescent coupling of two adjacent waveguides with a length of $>$700 ~$\mathrm{\mu m}$, provides wavelength-dependent optical transmission.
In this case, the WDM coupler is designed to provide unity transmission near 1550~nm and null transmission near 1480~nm, thus allowing for the combination/separation of 1480~nm optical pump and 1550~nm signal or laser critical to the emerging erbium-doped waveguide amplifiers ~\cite{liu2022photonic} and lasers~\cite{liu2023fully}.
Such WDM couplers fabricated using our approach show consistent performance over the stepper fields, while those using the photonic Damascene process show unpredictable variations of center wavelengths and transmission in optical transmission spectra (Fig.~\ref{fig:dispersion}(e)-(f)).
Moreover, we examined the performance of coupled-resonator systems ---critical to on-chip photonic dimmers~\cite{tikan2021emergent} and topological photonic chain~\cite{tusnin2023nonlinear}---that demand restricted control of waveguide dispersion and dimension.
Such coupled systems require the microring resonators to possess identical waveguide cross-section and circumference, to achieve system degeneracy.
Figure~\ref{fig:dispersion}(g) indicates that this desirable configuration can be achieved using our demonstrated subtractive process, showing two hybridized dispersion curves induced by two nearly identical resonators.
In comparison, it is not achievable for the photonic Damascene process due to the unavoidable variations of waveguide dimension and coupling gap even for two closely spaced microring resonators, which shows only a single dispersion curve in Fig.~\ref{fig:dispersion}(h).
Furthermore, we would like to note that the subtractive process is naturally suited for implementing more complicated devices such as arrayed waveguide gratings (AWG), which is however impossible for photonic Damascene process due to the waveguide dishing. 
Consequently, such comparisons were not conducted.

\subsection*{Microwave-band soliton microcomb generation}
\begin{figure*}
    \centering
    \includegraphics[width=1\textwidth]{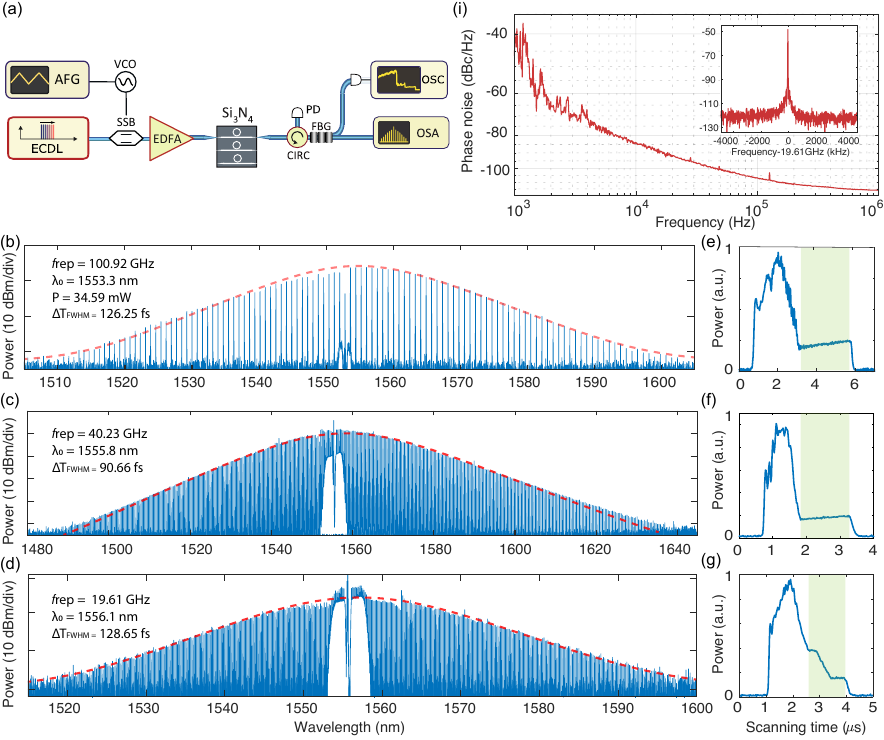}
    \caption{\textbf{Single soliton generation and the phase noise characterization of the soliton repetition rate.} \textbf{(a)} Experimental setup for soliton generation. A voltage-controlled oscillator (VCO) is utilized to generate a single-sideband (SSB) modulation for soliton seeding, with scanning speeds ranging from 100 kHz to 1 MHz. ECDL, external-cavity diode laser; VCO, voltage-controlled oscillator; EDFA, erbium-doped fiber amplifier; CIRC, circulator; FBG, fiber Bragg grating; PD, photodiode; OSC, oscilloscope; OSA, optical spectrum analyzer. \textbf{(b)(c)(d)} Single soliton spectra corresponding to repetition rates of 100.92~GHz, 40.23~GHz, and 19.61~GHz, respectively. \textbf{(e)(f)(g)} Soliton steps occurring in the resonator transmission spectrum during pump laser scanning across the resonance, aligning with the resonators featured in \textbf{(b)(c)(d)}.
    \textbf{h} Phase noise of a single soliton with 19.61~GHz repetition rate. The inset shows the beatnote of the soliton signal with a local oscillator for repetition rate measurement.}
    \label{fig:3}
\end{figure*}
To exemplify the dispersion tailoring and low loss achieved by the demonstrated fabrication approaches, we generated soliton microcombs at low threshold power levels using devices with a cross-section of 2.3~$\mu$m$\times$820~nm.
In Figure~\ref{fig:3}(b-d), we present the optical spectra of generated single-soliton state microcombs at repetition rates of 100.92~GHz, 40.23~GHz, and 19.61~GHz, respectively.
The corresponding characteristic soliton steps are illustrated in Fig.~\ref{fig:3}(e-g).
Low on-chip power levels are demonstrated for single soliton generation, specifically 34.59~mW for 100.92~GHz repetition rate single soliton, 111.98~mW for 40.23~GHz soliton, and 292.38~mW for 19.61~GHz soliton, attesting to the high-quality factor of our Si$_3$N$_4$ resonators.
%100 GHz: -5.8586 dB fiber to fiber
%40 GHz: -5.5516 dB fiber to fiber
%20 GHz: -7.2328 dB fiber to fiber
The single soliton spectrum has a 3-dB bandwidth of 20.15~nm with 25 comb lines at a repetition rate of 100.92~GHz. 
At 40.23~GHz, the bandwidth measures 28.1~nm with 86 comb lines, while at 19.61~GHz, the bandwidth is 19.77~nm with 125 comb lines. 
Figure \ref{fig:3}(i) illustrates the phase noise for the single soliton at 19.61~GHz repetition rate, measured with the self-heterodyne technique.
The inset presents the soliton's beatnote with a local oscillator, used for the determination of the repetition rate. 

\subsection*{UV irradiation-triggered enhancement of thermal absorption loss}

As waveguides are optimized to achieve ultra-low losses on the order of dB/m, thermal absorption emerges as a predominant factor contributing to the remaining loss.
The thermal absorption can impede the observation of the 'soliton steps' during slow laser scanning across the ring resonance, although its effect can be mitigated using fast laser scanning or dual-laser pumping techniques~\cite{zhou2019soliton}. 
Here, we report a decrease in thermal absorption loss in microresonators through a rapid thermal annealing (RTA) at 500~$^\circ$C for 1~hour.
This improvement is hypothesized to result from the reduction of defect centers in the Si$_3$N$_4$ layer induced by ultraviolet (UV) irradiation~\cite{neutens2018mitigation}.

\begin{figure*}
    \centering
    \includegraphics[width=1\textwidth]{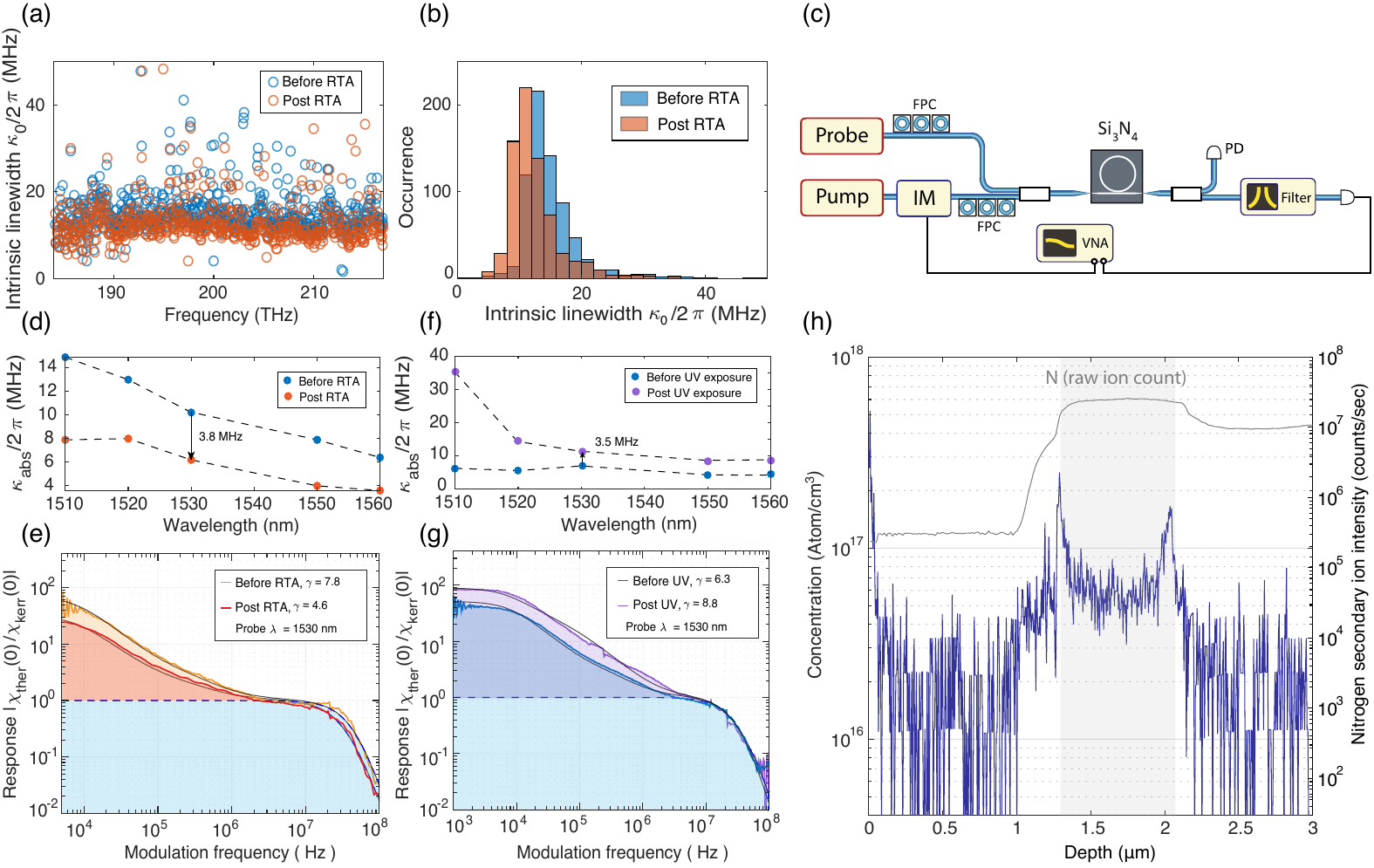}
    \caption{\textbf{Exploring the sources of thermal absorption in Si$_3$N$_4$ waveguides: UV-Induced defects, recovery via rapid annealing, and Cu impurities} \textbf{(a)} Intrinsic loss $\kappa_0/2\pi$ of fully SiO$_2$-cladded Si$_3$N$_4$ waveguides, pre and post rapid thermal annealing (RTA) at 500~$^\circ$C for 1~hour. \textbf{(b)} Histogram of $\kappa_0/2\pi$ in \textbf{(a)}. The mean, median, and most probable intrinsic loss rate $\kappa_0/2\pi$ shift from 15~MHz, 13.9~MHz, and 12~MHz to 12.6~MHz, 11.5~MHz, and 10~MHz, respectively. \textbf{(c)} Experimental setup of Kerr-nonlinearity-calibrated thermal response measurements. \textbf{(d)} Variations in fitted absorption $\kappa_{abs}/2\pi$ for devices before and after RTA. \textbf{(e)} A detailed response measurement at 1530~nm for pre and post RTA devices. \textbf{(f)} Variations in fitted absorption rates $\kappa_{abs}/2\pi$ for devices before and after undergoing UV irradiation in O$_2$ plasma stripper and UV tape curer. \textbf{(g)} A detailed response measurement at 1530~nm for pre- and post-UV irradiated devices. \textbf{(h)} Concentration profile of copper metal impurities in fully SiO$_2$-cladded Si$_3$N$_4$ waveguides. A correction factor based on Si$_3$N$_4$ area ratio is applied to the original data. The second ion intensity counts of nitrogen are plotted in grey for depth calibration.}
    \label{fig:4}
\end{figure*}

Optically active defect centers, known as K-centers (involving Si dangling bonds) and N-centers (with N dangling bonds), might be generated when Si-N bonds are broken by UV irradiation that has energy greater than the Si$_x$N$_y$ bandgap~\cite{warren1993si}.
These defect centers create specific electronic states, forming localized energy levels that can absorb and emit photons, thereby producing a detectable absorption signal. 
The interaction between UV light and Si$_3$N$_4$, however, is not solely confined to electronic transitions; it may also induce local heating or mechanical stress, which relaxes over time and causes changes in the atomic or molecular structure, leading to the formation of defects or imperfections within the lattice~\cite{miyagawa2007local,tian2010stress}.
The UV irradiation that triggers these effects may ubiquitously occur during various fabrication procedures, such as high-frequency plasma cleaning, UV ozone cleaning, or UV curing to remove carrier tape after wafer backside grinding.
To annihilate these defect states and initiate nitridation processes (formation of Si-N), we applied a rapid thermal annealing (RTA) to the fabricated device for 1 hour at 500~$^\circ$C in N$_2$ atmosphere.
The mean, median, and most probable intrinsic loss rates ($\kappa_0/2\pi$) decreased from 15~MHz, 13.9~MHz, and 12~MHz for as-fabricated devices to 12.6~MHz, 11.5~MHz, and 10~MHz for post-annealed devices, respectively (Fig.~\ref{fig:4}(a) and (b)).
To ascertain the origin of the loss change, we conducted broadband thermal absorption loss measurements~\cite{gao2022probing,liu2021high} on the devices before and after undergoing rapid thermal treatment.
A reduction in the fitted thermal absorption loss rate ($\kappa_{abs}/2\pi$, see "Methods") is observed within the 2.8-7~MHz range after applying RTA (Fig.~\ref{fig:4}(d)).
Fig.~\ref{fig:4}(e) illustrates the frequency responses, $\chi(\omega)=\chi_{\text{therm}}(\omega)+\chi_{\text{Kerr}}(\omega)$, at 1530~nm of pre- and post-RTA devices.
The fitted (black) and measured (red and yellow) responses are normalized to $\chi_{\text{Kerr}}$. 
The thermal and Kerr effects are identified through the fitting detailed in "Methods", and are delineated using yellow/red and blue colors, respectively.
The factor $\gamma = \frac{\chi_{\text{therm}}(0)}{\chi_{\text{Kerr}}(0)}$, proportional to
$\kappa_{\text{abs}}/2\pi$, decreases with post-RTA, indicating reduced thermal absorption.
%%%%%%%%%%%
To verify if the observed loss increase stems from UV-induced defect centers, we exposed post-RTA devices, identical in cross-section to those shown in Fig.~\ref{fig:4}(d)(e), to the same UV conditions used in the back end of fabrication process.
This involved a 600~W microwave oxygen plasma etcher for photoresist stripping and a 310~mW, 365~nm UV curer for tape removal after wafer backside grinding. 
Fig.~\ref{fig:4}(f)(g) depict the changes in $\kappa_{abs}/2\pi$ in the post-UV device, highlighting a 3.5~MHz loss increase at 1530~nm, comparable to the reduction seen in Fig.~\ref{fig:4}(d)(e).
From these findings, we confirm that the change in loss after RTA treatment is linked to the depopulation of the defect centers in Si$_3$N$_4$ generated by UV irradiation.

\subsection*{Copper-induced thermal absorption loss and concentration analysis}
Despite a decrease in UV-induced defect centers within the Si$_3$N$_4$ layer, the thermal absorption rate remains around 4-8~MHz, representing more than 50$\%$ of the total loss. 
Generally, Si$_3$N$_4$ is recognized as a transparent substance from the visible to the near-infrared with negligible absorption.
The additional loss can be attributed to impurities and defects within the lattice structure, such as vacancies, contaminants (e.g., oxygen or hydrogen), and interstitial atoms including transition metals.
Pfeiffer et al. assessed the concentration of various transition metal impurities in SiO$_2$ cladded Si$_3$N$_4$ waveguides created using the Damascene reflown process, identifying copper as a highly absorptive impurity, with a concentration of 10$^{18}$~atoms/cm$^{3}$ \cite{pfeiffer2018ultra}. 

%%%%%%%
Cu, along with Ni, Fe, and Cr, constitutes the most prevalent and detrimental metallic impurities in mc-Si and Si$_3$N$_4$, leading to the formation of absorption bands within the visible and near-infrared spectral domains~\cite{weber1983transition,beeler1990electronic}.
Specifically for copper, the Cu$^{2+}$ state exhibits strong absorption, with reported absorption per ppm impurity concentration ranging from 0.1 dB/km/ppm to several hundred dB/km/ppm, equating to an additional loss of $\kappa_0/2\pi$ = 10-100~MHz~\cite{newns1973absorption,ohishi1981impurity}.  
%Such findings underscore the particular concern regarding copper among metal impurities, as it demonstrates high solubility in both Si and Si$_3$N$_4$ compared to other elements, especially at elevated temperatures~\cite{istratov1998intrinsic,gilles1990impact,istratov2000diffusion,hozawa2004copper}.

To quantify the concentration of Cu impurities in our samples, we utilized Secondary Ion Mass Spectroscopy (SIMS) for localized probing of Cu concentration profile along the waveguide depth, with a total probing depth of 3~$\mu$m and a detection limit below 5$\times$10$^{15}$~atoms/cm$^3$. 
The secondary ion intensity for nitrogen is plotted in Fig.~\ref{fig:4}(h) in grey for Si$_3$N$_4$ waveguide layer determination and depth calibration.
Between depths of 0 and 1.1~$\mu$m, the Cu atom count remains at or below the detection limit, suggesting minimal to no Cu presence in SiO$_2$. 
For the analysis of the Si$_3$N$_4$ device layer below 1.1~$\mu$m, we examined a 20$\times$20~$\mu$m$^2$ area in the crater center, which consists of a 50-50 mixture of Si$_3$N$_4$ and SiO$_2$.
Since Cu detection arises solely from Si$_3$N$_4$ sputtering, the data is corrected with a factor of 2.
The quantified Cu content in Si$_3$N$_4$ waveguide layer is approximately 6$\times$10$^{16}$~atoms/cm$^3$.  
The saddle region at the Si$_3$N$_4$/SiO$_2$ boundary near 1.3~$\mu$m arises from a decrease in charge after SiO$_2$ etching and a complex ion count calibration due to different sputtering rates between SiO$_2$ and Si$_3$N$_4$.

\section*{Conclusion}

In summary, we demonstrated a novel DUV-based subtractive approach capable of producing high-density Si$_3$N$_4$ photonic integrated circuits with ultra-low propagation loss, precise control of dispersion and dimension, and compatibility with wafer-scale manufacturing.
We achieved, for the first time, the combination of these previously incompatible properties for reliable and consistent fabrication of Si$_3$N$_4$ photonic integrated circuits, eliminating the long-standing trade-off between low yield and uncontrollable dimension precision.
By implementing densely distributed, deeply etched connected filler patterns, we streamlined the deposition of thick LPCVD Si$_3$N$_4$ films into a single step, eliminating the need for thermal cycling to alleviate film stress while extending the storage time of the deposited film.
This approach not only sidesteps potential contamination arising from inter-layer particles but also drives down the manufacturing expenses of high-confinement Si$_3$N$_4$ PICs. 
%We also discovered the novel advantageous role of these deep trenches as effective copper impurity trapping centers that can diminish the absorption losses in Si$_3$N$_4$ waveguides.
Notably, we identified and implemented a post-fabrication rapid thermal annealing process as an effective countermeasure to losses induced by unavoidable UV exposure during fabrication.
With these collimated efforts, the fabricated microresonators achieved an average 24.17 million $Q$-value and the 0.5~m-long spiral waveguides exhibit losses within a 2.28-4~dB/m range in the optical C-band.
We demonstrated that our approach can achieve reliable control of dispersion, coupling strength, and dimension in individual resonators, WDM couplers, and coupled resonators, which will facilitate the future realization of large-scale Si$_3$N$_4$ photonic integrated circuits and networks.
We successfully generated Dissipative Kerr solitons in the Si$_3$N$_4$ resonators, achieving repetition rates as low as the K-band (19.61~GHz).
The process is cost-effective, scalable, and easily transferable to foundries, meeting future demands for large-scale, high-density photonic integrated circuits in communications, sensing, and quantum applications requiring on-chip linear and nonlinear optics.

\section*{Methods}

\subsection*{Analysis of thick LPCVD Si$_3$N$_4$ films with various growth methods}

A schematic comparison of our Si$_3$N$_4$ deposition with thermally cycled deposition is illustrated in Fig.~\ref{fig:0}(c)(d). 
A distinct boundary appears between the stacked films following intermediate annealing~\cite{ji2021methods,ye2019high} (Fig.~\ref{fig:0}(f)). 
Ye et al.~\cite{ye2019high} proposed that this boundary could be removed by a brief immersion in diluted hydrofluoric acid (HF) solution, indicating a possible link between the boundary and surface Si$_3$N$_4$ oxidation.
Notably, in the absence of intermediate annealing, this boundary is imperceptible~\cite{el2019ultralow,ye2023foundry}. 
Yet, when fabricating Si$_3$N$_4$ waveguides using such films, the etching endpoint detection (EDP) graph (Figure~\ref{fig:0}(g)) distinctly exhibits an endpoint midway through the process. 
This suggests that the cycled film lacks uniformity and potentially leads to increased scattering centers compared to the singly deposited film, even as the film boundary remains indiscernible.
An analogous trend extends to the etching endpoint of the cycled film subjected to intermediate annealing, revealing a broader midpoint with amplified magnitude (Figure~\ref{fig:0}(g)).
	
\subsection*{Probing the thermal absorption loss in Si$_3$N$_4$ waveguides}

In the experiment, we explored the resonance frequency shift response of a probe mode, induced by the intensity modulation of the pump mode, to calibrate the absorption loss of the pump resonance.  
Specifically, the pump laser is tuned to an optical resonance (frequency $\nu_p$) to characterize the thermal absorption loss $\kappa_{\text{abs}}/2\pi$, and is subsequently intensity-modulated at a frequency $\omega/2\pi$, swept from 1~kHz to 100~MHz. 
The modulation of the pump laser's intensity changes the photon number in the cavity, leading to a change in the refractive index via the optical Kerr effect.
The pump power is kept at an adequately low level, on the order of 10s~$\mu$W, to ensure that the steady-state frequency shift of the probe mode is minor relative to the resonance linewidth.
Simultaneously, the probe laser is loosely locked with an approximate 300~Hz locking bandwidth to a probe resonance, fixed near 1543~nm, within the filter transmission bandwidth of about 8~nm. 
The experimental setup is depicted in Fig.~\ref{fig:4}(c).
By leveraging the differing response times between the thermal and Kerr effects — with the thermal effect operating on a typical time scale of $10^{-3}$~s and the Kerr effect responding more rapidly — we discern between the two effects by carefully analyzing the frequency response of the probe laser's shifted resonance.
Subsequently, we extract the thermal-to-Kerr response ratio at zero modulation frequency and fit the thermal absorption rate of the Si$_3$N$_4$ waveguide using the following expression:
\begin{equation}
\kappa_{\mathrm{abs}}=\frac{2 c n_{\mathrm{mat}} n_2}{n_g n_{\mathrm{eff}} V_{\mathrm{eff}} \frac{\mathrm{d} T}{\mathrm{~d} P_{\mathrm{abs}}} \frac{\mathrm{d} n_{\mathrm{mat}}}{\mathrm{d} T}} \frac{\chi_{\text {therm }}(0)}{\chi_{\text {Kerr }}(0)}
\end{equation}
In this context, $V_\mathrm{eff}$ denotes the effective mode volume, $n_2 = 2.2\times10^{-19}$ m$^2$/W represents the nonlinear index of Si$_3$N$_4$~\cite{gao2022probing}. 
The group index is designated as $n_g = 2.1$, the material index as $n_{\text{mat}} = 2.0$, and the effective index as $n_{\mathrm{eff}}=1.8$. 
$\frac{dn}{dT} = 2.5 \times 10^{-5}$~K$^{-1}$ represents the thermo-optic coefficient of Si$_3$N$_4$, and $P_{\mathrm{abs}}$ indicates the absorbed power.
Finite-element simulations are performed to determine $V_{\mathrm{eff}}$ and to extract the frequency-domain thermal response function $\frac{\mathrm{d} T}{\mathrm{~d} P_{\mathrm{abs}}}$.
The factor $\gamma = \frac{\chi_{\text {therm }}(0)}{\chi_{\text {Kerr }}(0)}$ is the ratio between $\chi_{\text {therm }}(\omega)$ and $\chi_{\text {Kerr }}(\omega)$ (derived from fitting the response function $\chi(\omega)=\chi_{\text {therm }}(\omega)+\chi_{\text {Kerr }}(\omega)$), evaluated at zero modulation frequency.
The derivation of this expression can be found in~\cite{liu2021high}.
The measurement results are available in Supplementary Information Note 3.

\subsection*{Soliton generation with single sideband fast scanning}
	
In this section, we detail the soliton microcomb pumping approach using a high-speed single-sideband (SSB) scan. 
Thermal absorption in microresonators can introduce large resonance frequency drifts when transitioning from modulation instability to the soliton regime, complicating soliton step observation during frequency sweeps.
To circumvent this, we leveraged a modulation strategy involving two phase modulators and an amplitude modulator to suppress the primary laser carrier and generate a single optical sideband, achieving a pump mode suppression ratio beyond 20~dB.
The modulated laser, now carrying the novel optical sideband, is directed into the microresonator.
This enables high-speed scans ranging from 100 kHz to 1 MHz, governed by a voltage-controlled oscillator (VCO).
This method circumvents thermal absorption — an impediment due to its comparable timescale with direct laser tuning speed — by leveraging the superior scanning rate of the SSB, which outpaces the millisecond-level thermal response, enabling the emergence of soliton steps.
In Supplementary Information Note 6, we benchmark the frequency noise of the single sideband with the unmodulated continuous-wave laser. 
Notably, both exhibit comparable noise profiles, with additional peaks proximate to the modulation frequency in single sideband scanning.

%\noindent \textbf{Methods}
\medskip
\begin{footnotesize}
\noindent \textbf{Funding Information}: This work was supported by the Air Force Office of Scientific Research (AFOSR) under Award No. FA9550-19-1-0250, by the EU H2020 research and innovation program under grant No. 965124 (FEMTOCHIP), 101017237 (PHOENICS), and 863322 (TeraSlice). 
Y.L. acknowledges support from Swiss National Science Foundation under grant No. 221540 (Bridge PoC). 
This work was also supported by the EU Horizon Europe EIC programme under grant agreement No. 101047289 (CSOC) and funded by the Swiss State Secretariat for Education, Research and lnnovation (SERI).

\noindent \textbf{Acknowledgments}: 
Silicon nitride samples were fabricated in the EPFL Center of MicroNanoTechnology (CMi).  

\noindent \textbf{Author contributions}: 
X.J. simulated and designed the devices.
X.J. fabricated the device, with the help of R.N.W..
X.J. characterized the devices, with the help of J.R.. 
X.J. did the experiments and analyzed the data with the help of Y.L. and J.R. 
X.J. wrote the manuscript, with input from all authors. 
T.J.K. supervised the project.

\noindent \textbf{Data Availability}: 
The code and data used to produce the plots within this work will be released on the repository \texttt{Zenodo} upon publication of this preprint.

\noindent \textbf{Competing interests}
T.J.K. is a cofounder and shareholder of LiGenTec SA, a start-up company offering \ce{Si3N4} photonic integrated circuits as a foundry service. 
The other authors declare no competing interests.

\end{footnotesize}
%\vspace{-0.3cm}

\renewcommand{\bibpreamble}{
$\dag$\textcolor{magenta}{tobias.kippenberg@epfl.ch}
}
\pretolerance=0
\bigskip
%\bibliographystyle{apsrev4-2}
%\bibliography{zotero_updated_v1, library_additional}

\bibliography{biblio}

\end{document}

%\includepdf[landscape=false]{Science_SM_Cover.pdf}
\title{Supplementary Information for: Foundry compatible, efficient wafer-scale manufacturing of ultra-low loss, high-density Si$_3$N$_4$ photonic integrated circuits}

\author{Xinru Ji$^{1,2}$, Rui Ning Wang$^{1,2}$, Yang Liu$^{1,2}$, Johann Riemensberger$^{1,2}$, Zheru Qiu$^{1,2}$, and Tobias J. Kippenberg$^{1,2,\ddag}$}

\affiliation{
$^1$Institute of Physics, Swiss Federal Institute of Technology Lausanne (EPFL), CH-1015 Lausanne, Switzerland\\
$^2$Center for Quantum Science and Engineering,Swiss Federal Institute of Technology Lausanne (EPFL), CH-1015 Lausanne, Switzerland
}

%%%%%% RESET EQUATION NUMBERS ETC. %%%%%%%%
\setcounter{equation}{0}
\setcounter{figure}{0}
\setcounter{table}{0}

\setcounter{subsection}{0}
\setcounter{section}{0}
\setcounter{secnumdepth}{3}

 \begin{abstract}
 Supplementary Information accompanying the manuscript containing long-term storage of thick LPCVD Si$_3$N$_4$ films, annealing conditions, thermal absorption loss measurement, scattering loss analysis, and experimental details for soliton generation.
 \end{abstract}

\maketitle
{\hypersetup{linkcolor=blue}\tableofcontents}
\newpage

%%%%%%%%%%%%%%%%%%%%%%%%%%%%%%%%%%%%%%%%%%%%%%%%%%%%%%%%%%%%%%%
\section{Long-term storage of LPCVD Si$_3$N$_4$ films enabled by deeply etched filler patterns}
This section focuses on the prolonged storage of LPCVD Si$_3$N$_4$ films, which is made possible through the utilization of densely distributed filler patterns.
We applied LPCVD Si$_3$N$_4$ films with a thickness of 820~nm on 4-inch Si wafers in three different configurations: Wafer 1, which is a patterned Si wafer with a 4 $\mu$m thermal oxide layer; Wafer 2, which is a regular Si wafer with a 4 $\mu$m thermal oxide film; and Wafer 3, which is a pure Si wafer.

The Si$_3$N$_4$ layer deposited using the LPCVD technique was applied on both surfaces of the wafer to mitigate tensile stress and minimize the occurrence of wafer bowing.
Normally, the initiation of cracks occurs at the periphery of the wafer during high-temperature treatments.
This phenomenon can be attributed to the different thermal expansion coefficients exhibited by the films and the silicon substrate.
To examine the process of crack formation and assess the efficacy of our filler patterns, wafer 1 incorporates a deliberate omission of patterns at its edge.  

\begin{figure*}[htb]
    \centering
    \includegraphics[width=1\textwidth]{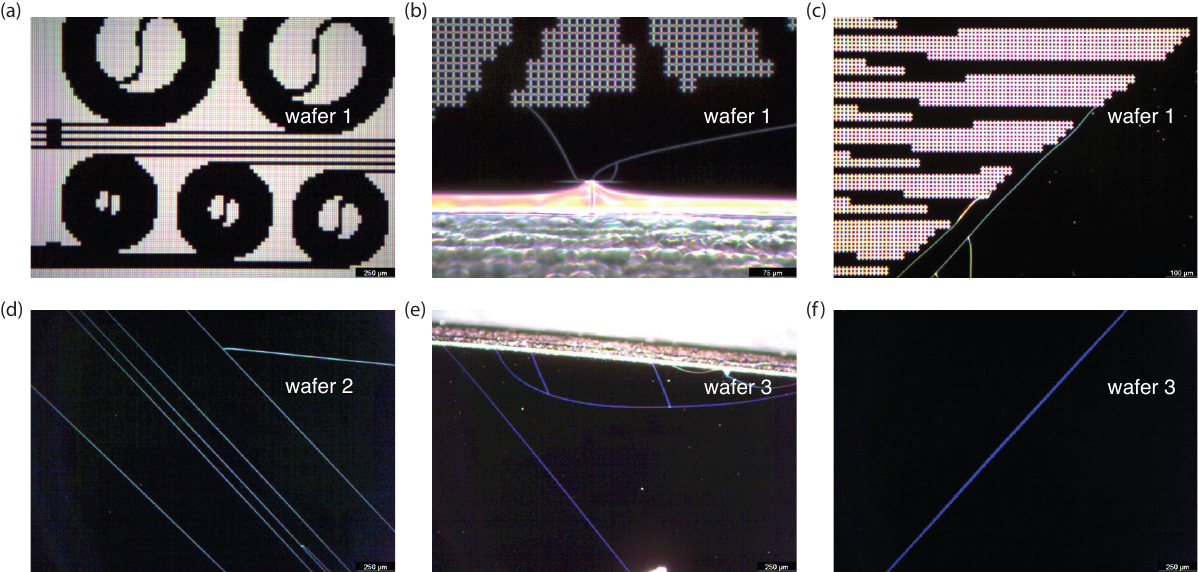}
    \caption{\textbf{Storage assessment of 820~nm-thick LPCVD Si$_3$N$_4$ films on varied 4-inch Si wafers.} 
    \textbf{(a)} A dark field optical microscopic image of 820~nm-thick LPCVD Si$_3$N$_4$ on patterned Si with 4~$\mu$m SiO$_2$ (annotated as wafer 1). 
    Black regions, reserved for spiral waveguides (0.32~mm$^2$ to 1.57~mm$^2$ in footprint), show no cracks.
    \textbf{(b)} Cracks at wafer edge, revealing origins due to the difference in thermal expansion between Si$_3$N$_4$ and SiO$_2$ during high-temperature treatment.
    \textbf{(c)} Edge-originating cracks halted by filler pattern trenches during propagation.
    \textbf{(d)} Cracks in 820~nm-thick LPCVD Si$_3$N$_4$ on a Si wafer with 4~$\mu$m SiO$_2$ (wafer 2). 
    \textbf{(e)(f)} Edge and central cracks of 820~nm-thick LPCVD Si$_3$N$_4$ film on a Si wafer (wafer 3).} 
    \label{SI:fig:1}
\end{figure*}

After a storage period of 5 months, it was observed that wafer 1 did not exhibit any crack in the center of the wafer (Supplementary Figure \ref{SI:fig:1}(a)), even in regions designated for 0.5-m long spiral waveguides of 4200$\times$1050~$\mu$m$^2$ in footprint. 
This is due to the efficient interruption of crack propagation by our interconnected filler pattern (Supplementary Figure \ref{SI:fig:1}(c)). 
On the other hand, wafers 2 and 3 have discernible cracks beginning from the edges, which propagate over the entire wafer, significantly impacting the operation of the device and the overall yield of the fabrication process. 
The ability to store Si$_3$N$_4$ films for an extended period eliminates the need for quick post-deposition processing, optimizing the LPCVD furnace's deposition capacity and decreasing costs for high confinement Si$_3$N$_4$ platforms.

\section{Impact of different annealing conditions for LPCVD Si$_3$N$_4$ films}
\begin{figure*}[htb]
    \centering
    \includegraphics[width=1\textwidth]{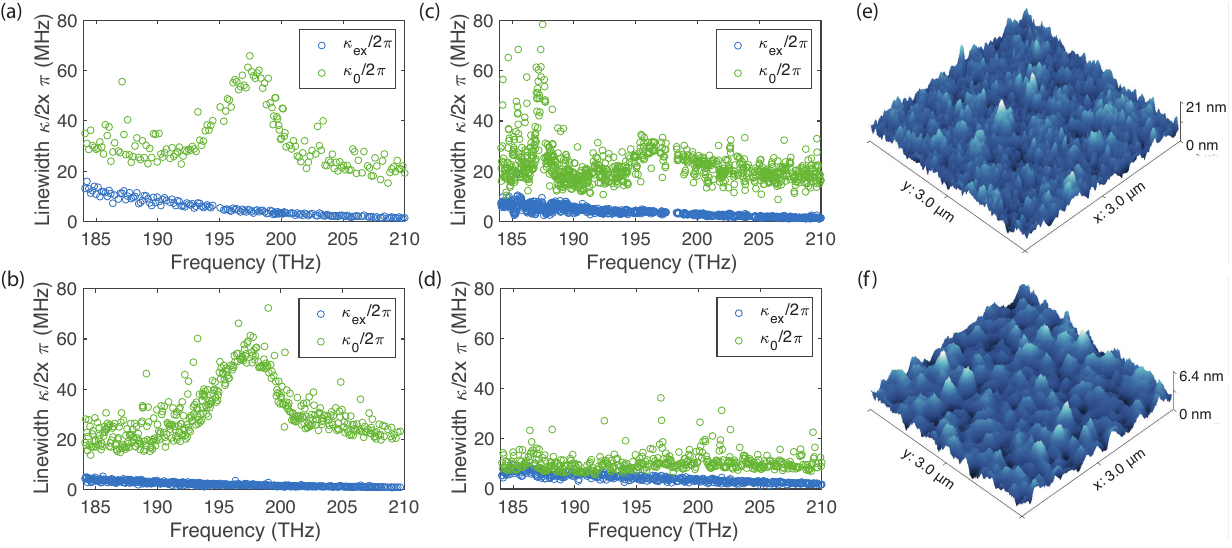}
    \caption{\textbf{Impact of thermal annealing conditions on the loss of Si$_3$N$_4$ waveguides.} 
    \textbf{(a)} Resonator loss after 9~hours annealing at 1050~$^\circ \text{C}$ for Si$_3$N$_4$ and 6~hours annealing at 900~$^\circ \text{C}$ for SiO$_2$.
    \textbf{(b)} Resonator loss after 3~hours annealing at 1200~$^\circ \text{C}$ for Si$_3$N$_4$ and 11~hours annealing at 1200~$^\circ \text{C}$ for SiO$_2$.
    \textbf{(c)} Resonator loss processed with 6 hours annealing for Si$_3$N$_4$ at 1200~$^\circ \text{C}$ and 11 hours annealing for SiO$_2$ at 1200~$^\circ \text{C}$. 
    \textbf{(d)} Resonator loss processed with 11 hours annealing for Si$_3$N$_4$ at 1200~$^\circ \text{C}$ and 11 hours annealing for SiO$_2$ at 1200~$^\circ \text{C}$. 
    \textbf{(e)} Atomic force microscopy (AFM) scan of as-deposited 1.1~$\mu$m SiO$_2$ surface, displaying a RMS roughness of 2.29nm.
    \textbf{(f)} AFM scan of annealed 1.1$\mu$m SiO$_2$ surface (1200~$^\circ \text{C}$, 11 hours), exhibiting a reduced RMS roughness of 0.76~nm.}
    \label{SI:fig:2}
\end{figure*}

During the deposition of the Si$_3$N$_4$ layer, the incorporation of hydrogenated precursors (SiH$_2$Cl$_2$ and NH$_3$) leads to the formation of numerous N-H and Si-H bonds in the as-deposited Si$_3$N$_4$, resulting in intrinsic infrared absorption~\cite{germann2000silicon,lemaire1986optical,pfeiffer2018ultra}. 
The reduction of residual hydrogen content has been seen during elevated-temperature annealing of both the Si$_3$N$_4$ core and the SiO$_2$ cladding films~\cite{boehme2000h}.
This release of hydrogen could occur via "slow" atomic diffusion involving covalently bonded atoms migrating between binding sites, or through "fast" molecular diffusion of hydrogen-containing molecules such as H$_2$, NH$_3$, and HCl, which dissociate from the network prior to diffusion. 
Consequently, identifying optimal annealing conditions becomes crucial to circumvent additional absorption losses in Si$_3$N$_4$ waveguides. 
To determine the suitable annealing condition, we implement 4 protocols, summarized in Supplementary table \ref{SI:tab:1}.

\begin{table}[h]
\centering
\caption{\bf Four annealing protocols evaluated for optimal conditions}
\begin{tabular}{lcccc}
\hline
\textbf{layer/scheme} & \textbf{1} & \textbf{2} & \textbf{3} & \textbf{4} \\
\hline
\textbf{Si$_3$N$_4$} & 1050$^\circ \text{C}$ $\times$ 9 hours & 1200$^\circ \text{C}$ $\times$ 3 hours & 1200$^\circ \text{C}$ $\times$ 6 hours & 1200$^\circ \text{C}$ $\times$ 11 hours \\
\textbf{SiO$_2$} & 900$^\circ \text{C}$ $\times$ 6 hours & 1200$^\circ \text{C}$ $\times$ 11 hours & 1200$^\circ \text{C}$ $\times$ 11 hours & 1200$^\circ \text{C}$ $\times$ 11 hours \\
\hline
\end{tabular}
  \label{SI:tab:1}
\end{table}

The linear characterization of resonators, subjected to the four annealing protocols, is presented in Supplementary figure \ref{SI:fig:2}. 
These resonators have a consistent 50~GHz FSR but differ in cross-sections, yet all within the anomalous dispersion regime. 
Rather than comparing intrinsic loss values, our focus here lies on absorption characteristics. 

Schemes 1, 2, and 3 exhibit a broad overtone near 197~THz, a combination of N-H absorption at 199~THz, and Si-H at 195~THz, resulting in an overlapped peak evident from linewidth measurements. 
The tail of this absorption peak imposes additional losses in the 1427~nm to 1620~nm wavelength range. 
Notably, a tail around 185~THz to 190~THz for sample 1 and a peak near 187~THz in sample 3 (Supplementary figure \ref{SI:fig:2}(a)(c)) can be further attributed to water absorption~\cite{afrin2013water}—a consequence of insufficient thermal treatment. 
Extended annealing at 1200~$^\circ \text{C}$ for 11 hours effectively reduces N-H and Si-H bonds and residual water in Si$_3$N$_4$ and SiO$_2$ layers (scheme 4), evident from the flat spectral dependence of intrinsic loss in sample 4 (Supplementary figure \ref{SI:fig:2}(e)). 
The results underscore the importance of providing sufficient thermal excitation energy to overcome the binding energy between hydrogen and nitrogen or silicon atoms. 
Additionally, high-temperature treatment improves the SiO$_2$ film's surface RMS roughness, reducing from 2.29~nm to 0.76~nm (Supplementary figure \ref{SI:fig:2}(e)(f)), attributed to silicon dioxide molecular reconfiguration and void reduction during deposition.

\section{Quantitative evaluation of thermal absorption loss through linear response measurement}

\begin{figure*}[h!]
    \centering
    \includegraphics[width=1\textwidth]{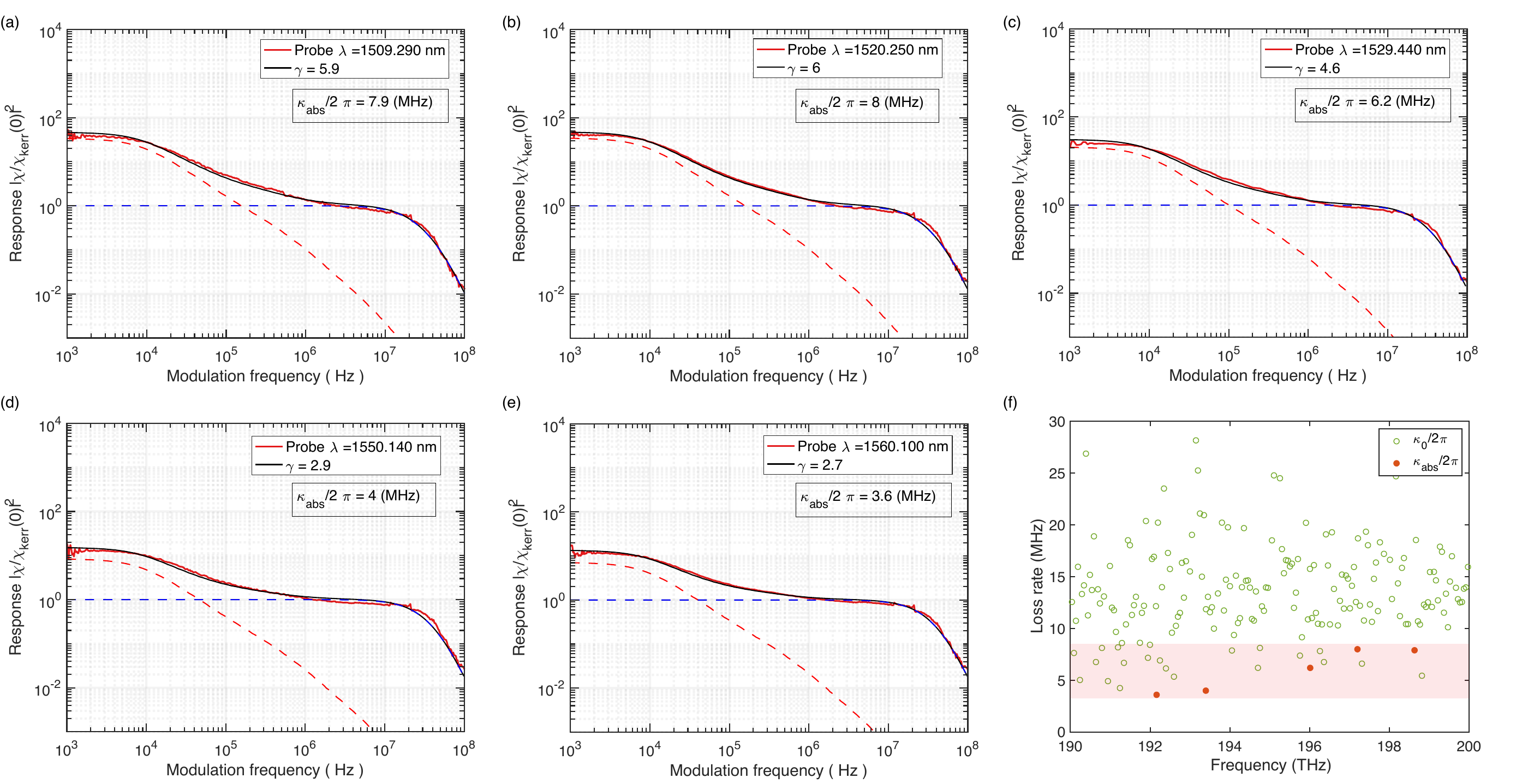}
    \caption{\textbf{Calibrated absorption loss} $\boldsymbol{\kappa_{abs}/2\pi}$ \textbf{of different resonances in a 50~GHz microresonator.}
    The microresonator has a cross-section of 720$\times$2300~nm$^2$.
    The measured frequency response $\chi (\omega)$ in red (fitting in black), is normalized to $\chi_{\text{Kerr}}$.
    Five response measurements at \textbf{(a)} 1509.29~nm, \textbf{(b)} 1520.25~nm, \textbf{(c)} 1529.44~nm, \textbf{(d)} 1550.14~nm, \textbf{(e)} 1560.1~nm, are presented. 
    \textbf{(f)} depicts intrinsic loss rate ($\kappa_{0}/2\pi$) and absorption loss rate ($\kappa_{abs}/2\pi$) over the photonic C-band, highlighting absorption's contribution to total loss.
}
    \label{SI:fig:4}
\end{figure*}

This section focuses on the quantitative assessment of thermal absorption loss through linear response measurement described in the main text. 
The resonator under assessment has a cross-section of 720$\times$2300~nm$^2$, and is post-annealed at 500$^\circ \text{C}$ for 1 hour to eliminate UV-induced absorption.
The pump laser is tuned to optical resonances ($\lambda_p$) proximate to 1510~nm, 1520~nm, 1530~nm, 1550~nm, and 1560~nm respectively. 
Concurrently, intensity modulation of the pump laser is performed at a frequency of $\omega$/2$\pi$, systematically swept across a range from 1~kHz to 100~MHz.
The resonance frequency shift responses of the probe mode at the corresponding pump laser wavelength are shown in SI Fig.~\ref{SI:fig:4}(a)-(e).
SI Fig.~\ref{SI:fig:4}(f) illustrates the thermal absorption loss $\kappa_{abs}$/2$\pi$ and its significant contribution, exceeding 40$\%$, to the overall resonator loss.

\section{Scattering loss simulation}

\begin{figure*}[h!]
	\centering
	\includegraphics[width=0.9\textwidth]{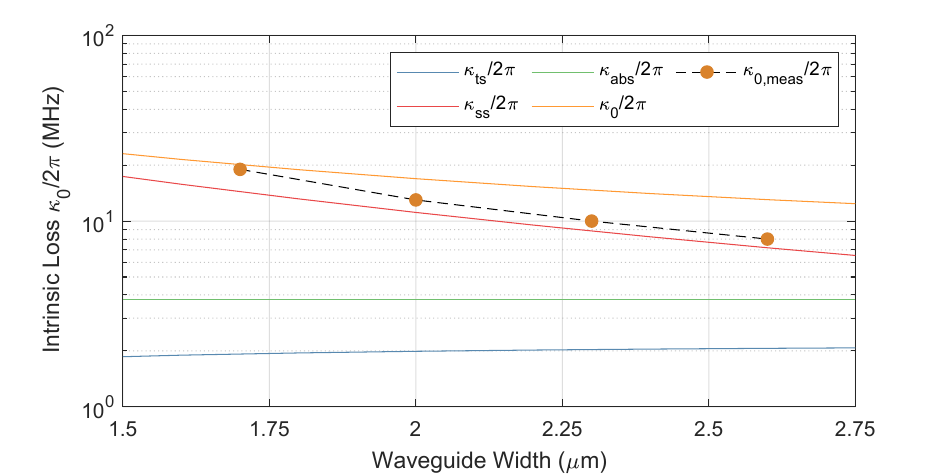}
	\caption{\textbf{Top and side wall scattering loss model.} Comparison of measured intrinsic cavity loss $\kappa_\mathrm{0,meas}/2\pi$ as a function of the waveguide width (orange markers). The simulated microresonator loss (orange line) is composed of the side wall scattering loss (red), the top and bottom wall scattering loss (green), and the measured absorption loss (blue).
	}
	\label{SI:fig:Xinru_Measured_Losses_MeasAbs}
\end{figure*}

The total intrinsic loss $\kappa_{tot}$ in optical microresonators is composed of multiple contributions of scattering and absorption:
\begin{equation}
	\kappa_{0} = \kappa_{abs} + \kappa_{ss} + \kappa_{ts} + \kappa_{bs}
\end{equation}
wherein $\kappa_{abs}$ denotes the optical absorption loss that has been measured using thermal absorption spectroscopy, $\kappa_{ss}$ denotes the scattering at the waveguide sidewall, and $\kappa_{ts}$ ($\kappa_{bs}$) denotes the scattering at the top (bottom) surfaces of the waveguide. 
To further disentangle the origin of waveguide losses in our structures, we numerically simulated the loss contribution of the light scattering from sidewall imperfections based on the equivalent current model proposed by Lacey and Payne \cite{payne1994theoretical}: 
\begin{equation}
	\kappa_{ts,ss} = \dfrac{n}{} \phi^2 \cdot \left(n_\mathrm{eff}^2-n_\mathrm{clad}^2\right) \cdot \frac{k_0^3}{4\pi n_\mathrm{eff}} \cdot S 
\end{equation}
with the normalized field integrated along the scattering interface $ \phi = \int E(x,y) dl / \iint E(x,y) dA$. 
The modal field is determined using a finite element solver in a bent waveguide with a radius of 231~$\mu$m, corresponding to a 100~GHz optical microring resonator.
The scattering integral $S$ is calculated from the exponential autocorrelation function of the sidewall roughness profile $R(u) = \sigma^2 \exp{-u^2/L_\textrm{c}^2}$, where $\sigma$ denotes the RMS roughness of the sidewall and $L_\mathrm{c}$ denotes the correlation length:
\begin{equation}
	S = \sqrt{2}\sigma^2 L_\mathrm{c}\pi \sqrt{\dfrac{\sqrt{M^2 + N^2} + N}{M^2 + N^2}}
\end{equation}
with the parameters $M = 2\beta L_\mathrm{c}$ and $N = 1 - L_\mathrm{c}^2 \left( \beta^2 - n_\mathrm{clad}^2 k_0^2 \right)$.
Light scattering occurs both at the top and bottom surfaces and sidewalls of the waveguide. 
The top surface roughness was measured with atomic force microscopy to be 0.25~nm with an exponential roughness power spectral density and a roughness correlation length of 13~nm, contributing less than 0.5~MHz to the intrinsic cavity loss rate and the absorption loss is determined as 4~MHz. 
The horizontal sidewalls contribute much more strongly to the intrinsic cavity losses. 
The overlap of the electric field with the sidewall strongly depends on the width of the optical waveguide, while the top overlap is only weakly affected by the waveguide width. 
We used SEM with 80~k magnification to determine the side wall line edge roughness, for which we found an upper limit of the RMS roughness of between 1~nm and 1.5~nm limited by the resolution of the SEM of 1.33~nm per pixel and a roughness correlation length of 50-100~nm. 
Indeed, we find that the measured loss values are consistent with an RMS sidewall roughness value of $\sigma_\mathrm{ss} = 0.75~nm$ (cf.~Fig.~\ref{SI:fig:Xinru_Measured_Losses_MeasAbs}). 
It should also be noted that the 2D roughness model is known to overestimate the scattering loss of rectangular optical waveguides by a moderate amount depending on the waveguide height \cite{barwicz2005three}. 

\section{Soliton microcomb pumping with single-sideband fast scanning}

\begin{figure*}[h!]
    \centering
    \includegraphics[width=0.8\textwidth]{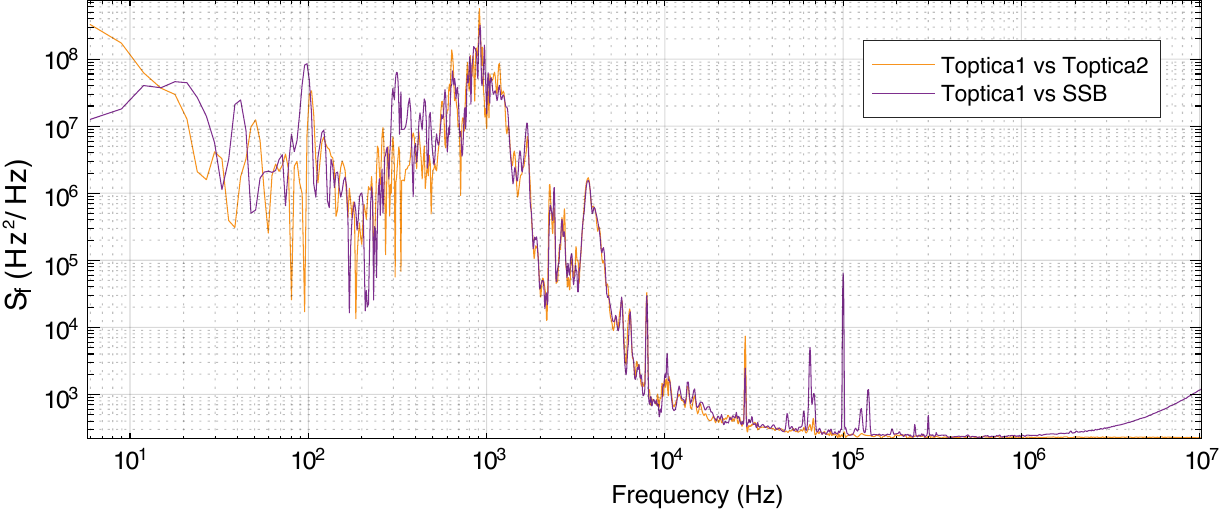}
    \caption{\textbf{Frequency noise comparison between the generated single optical sideband and unmodulated Toptica CTL diode laser.}
    A single optical sideband is generated using two phase modulators and one amplitude modulator, modulated at 100~kHz. 
    Noise characteristics are assessed via heterodyne beat note measurement using another Toptica CTL diode laser as the local oscillator.
    Both the single sideband and the unmodulated laser exhibit similar noise performance, with a few distinct peaks near the modulation frequency.}
    \label{SI:fig:3}
\end{figure*}

We benchmarked the frequency noise from a Toptica CTL diode laser against that from the VCO-driven SSB, utilizing a second Toptica CTL as a reference oscillator.
The frequency noise of SSB, presented in SI Fig.~\ref{SI:fig:3}, depicts a performance akin to the standalone Toptica laser, except for a few peaks around 100~kHz VCO modulation frequency. 
Although an even more stable frequency source would enhance the noise performance comparison, our findings suggest that the SSB-driven system introduces negligible noise compared to the direct laser pumping scheme for soliton generation.

\clearpage
%%%%%%%%%%%%%%%%%%%%%%%%%%%%%%%%%%%%%%%%%%%%%%%%%%%%%%%%%%%%%%%%

\bigskip
\bibliographystyle{apsrev4-2}
\bibliography{ref_si}